\documentstyle{article}
\def\be{ \begin{equation}}
\def\ee{ \end{equation}}
\def\bea{ \begin{eqnarray}}
\def\eea{ \end{eqnarray}}
\newtheorem{theo}{Theorem}
\newtheorem{defn}[theo]{Definition}

\title{Models of Quantum Space Time:  Quantum Field Planes }
\author{{\sc Gerhard Mack}
                      \\
        II. Institut f\"ur Theoretische Physik,
        Universit\"at Hamburg, Germany
        \\[2mm]
        {\sc Volker Schomerus} \\
    Harvard University, Cambridge, MA 02138, U.S.A.
    \thanks{Partly supported by NSF Grant No. PHY-91-20626}}
\date{}

\def\ba{\bea}
\def\ea{\eea}
\def\1{\iota }

\def\a{\alpha }
\def\b{\beta }
\def\c{\gamma }
\def\k{\kappa }

\def\k{\kappa }
\newcommand{\vvert}[3]{(_{#1\ #2}^{\ \ #3})}
\def\Rep{{\cal R}{\it ep}}
\def\<>{\stackrel{\scriptscriptstyle < }{ \scriptscriptstyle >}}
\def\1{\iota }
\def\O{{\cal O}}
\def\T{{\cal T}}
\def\R{{\cal R}}
\def\D{\Delta }
\def\H{{\cal H}}

\def\A{{\cal A}}

\def\G{{\cal G}}
\def\F{{\cal F}}

\def\C{{\cal C}}

\def\K{{\cal K}}

\def\t{\tau}

\def\vp{\varphi }
\def\s{\sigma }
\def\Th{\Theta}

\def\End{{\mbox{\it End\/}}}
\def\bmu{{\bar \mu}}
\def\brho{{\bar \rho}}
\def\pl{\partial}
\def\nn{\nonumber}

\def\e{\epsilon}
\def\S{{\cal S}}
\def\o{\otimes }
\def\bo{\mbox{\,\raisebox{-0.65mm}{$\Box$} \hspace{-4.7mm}
${\scriptstyle\times}$ \/}}

\def\Om{\Omega}
\begin{document}
\maketitle
\begin{abstract}
Quantum field planes furnish a noncommutative
differential algebra $\Om$ which
substitutes for the commutative algebra of functions and forms on
a contractible manifold. The data requirered in their
construction come from a quantum field theory. The basic idea is
to replace the ground field ${\bf C}$ of quantum planes
by the noncommutative algebra $\A$ of observables
of the quantum field theory.
\end{abstract}
\vspace*{-9.5cm}
\hspace*{10.7cm}
{\large \tt HUTMP 94-B335} \\
\hspace*{10.7cm}
{\large \tt hep-th/9403170} \\[9cm]

\section{A picture of quantum space time}
\def\tt{\tilde \tau}
\setcounter{equation}{0}

We wish to compound a global space time, which is quantum, from
local Lorentz frames. The physics in the local Lorentz frames shall be
described by a special relativistic  local quantum
field theory on a classical space time of some dimension.

To accomplish such a construction,
we need to generalize the classical setup of
general relativity which involves a classical space time manifold
and a differential calculus and geometry on it \cite{Con}.

Quantum field planes furnish a noncommutative
differential algebra $\Om$ which
substitutes for the commutative algebra of functions and forms on
a contractible manifold. The basic data to construct it come from
a  quantum  field theory in the local Lorentz frame.
\be \Om = \F \o \Phi \ee
as a linear space, where $\F $ plays the role of an "algebra of
scalar functions on quantum space time", and the fibre $\Phi $ is
the algebra of field operators of the quantum field theory
in the local Lorentz frame.

Quantum  field planes are generalizations of the quantum planes which
were studied by Manin \cite{Manin}, Wess and Zumino \cite{WeZu},
and others,
and were generalized to the quasi-associative case by he authors
\cite{MSVI}. Basically the
construction of quantum field planes
replaces the ground field ${\bf C}$ of quantum planes
by the noncommutative
algebra $\A $ of observables of the quantum field
theory in the local
Lorentz frame. Its elements will be  called {\it $a$-numbers\/}.

There is one extra piece of structure which
has no analogue in quantum planes. It comes
from the operator product expansions in the  quantum field theory
$\Phi $.

We begin with the general description of the construction of the
differential calculus, without imposing
operator product expansions at first. Then examples will be presented,
including the classical calculus, quantum planes, and
quantum field planes. Finally, the extra consistency
conditions from operator product expansions will be handled.

The Doplicher Haag Roberts construction of superselection sectors
in local quantum field theories is basic \cite{DHR1,DHR2}. It starts
from localizable morphisms of the
algebra of observables. The reconstruction theorem of
Doplicher and Roberts for 4-dimensional theories \cite{DoRo}
was recently generalized by one of us to
situations with braid statistics \cite{VSDiss}.
Here we will use the language of amplimorphisms which was introduced
be Vecsernyes and Szlachanyi\cite{VecSzl}. It simplifies our proofs
considerably and permits a unified point of view. Moreover,
quasi-associative possibilities appear very natural in this
setting.

\section{Differential calculus on generalized quantum planes}
        \label{DC}
\setcounter{equation}{0}

In the first subsection we describe an algebraic structure,
which is often met in physics. We will show later that
quasi-Hopf algebras as well as observable algebras in
quantum field theory provide examples. This algebraic
structure is sufficient for many constructions known in
the theory of quantum groups. As an example we will discuss
how quantum planes and differential calculi thereon can
be obtained within the general framework.

\subsection{Amplimorphisms and their intertwiners}

Throughout this paper, let \\[2mm]
$\K $ - *-algebra with unit e . \\
$\Rep$ - a set of *-representations $\pi$ of $\K$. \\
$\pi_0$ - a fixed representation in $\Rep$. It will act
   in a Hilbert space  $\H_0$               \\[2mm]
Let $W$ be some finite dimensional Hilbert space.
\begin{defn}{\em \cite{VecSzl,Vec} (Amplimorphisms) }
An amplimorphism
$$ \mu    : \K \mapsto
                \End   (W ) \o \K$$
is a not necessarily unit-preserving *-homomorphism of algebras.
\end{defn}
Given the  representation $\pi_0 $ of $\K $ and an amplimorphism
$\mu : \K \mapsto \End (W)\o \K $, a new representation
\be \pi = \pi_0\circ \mu \equiv (id \o \pi_0)\circ\mu .\label{muRep} \ee
can be constructed. Its representation space is $ W\o \H_0$.

This construction principle is very powerful. It
  generalizes the use of morphisms in
the theory of superselection sectors of algebraic field theory.
Using the fact that amplimorphisms can be
composed in a natural way,
$$ \mu_2 \circ \mu_1: \ \K  \mapsto \End (W_1)\o \End (W_2)\o \K , $$
they lead to a definition of a product of representations
of $\K $.
\begin{defn}
{\em (product of representations of $\K $)}
Suppose that $\pi_i=\pi_0 \circ \mu_i $. Then
 $$  (\pi_2 \bo \pi_1)(\xi) \equiv \pi_0
     ((\mu_{1} \circ \mu_{2})(\xi)).$$
\end{defn}
Now let us consider a family of Hilbert spaces,
$ W_\pi, \pi \in \Rep , \  W_{\pi_0} = {\bf C}$
and a family of amplimorphisms
$\mu_\pi, \pi \in \Rep$  such that
\begin{enumerate}
\item          $\mu_{\pi_0} = id_\K$,
\item $ \pi_0 \circ \mu_{\pi_1} \circ \dots \circ \mu_{\pi_n} \in \Rep$,
\item $ \pi(\xi) = \pi_0 \circ \mu_\pi (\xi)$.
\end{enumerate}
%
\begin{defn}\label{Fam} A family of amplimorphisms with the
properties {\it 1.,2.,3. } is said to generate all the
representations $\pi\in \Rep $ of $\K $ from $\pi_0$.
\end{defn}
Besides the amplimorphisms, one considers their intertwiners.
\begin{defn}
{\em (intertwiners)}
Given amplimorphisms $\mu_i: \K \mapsto \End (W_i) \o \K$,
an element                                                 $T \in
     \End (W_1,W_2) \o \K$ is called an intertwiner
between $\mu_1 $ and $\mu_2 $ if
   $$ T \mu_1(\xi) = \mu_2 (\xi)T $$
\end{defn}
Of particular interest are intertwiners
 $ \vp(\mu_\pi, \mu_\s) \in \End(W_{\s \bo \pi},
     W_\s \o W_\pi) \o \K $
which intertwine $\mu_\pi \circ \mu_\s $ and $\mu_{\s \bo \pi }$, viz.
\be  \vp(\mu_\pi, \mu_\s) \mu_{\s \bo \pi} (\xi) =
     \mu_\pi \circ \mu_\s (\xi) \vp( \mu_\pi, \mu_\s) \label{qass}\ee
Note the contravariant nature of this relation: The order of
$\s $ and $\pi $ gets interchanged.

An assignment $\vp :\, (\mu_\pi, \mu_\s ) \mapsto
               \vp(\mu_\pi, \mu_\s ) $
with this intertwining property
will be called a {\it reassociator\/}. The name comes from the fact
that the product $\circ $ of amplimorphisms is always associative,
while the product $\bo $ of representations of $\K $ is not necessarily
associative, but only quasiassociative in the sense that
a relation of the form (\ref{qass})
holds.

Statistics elements are another important class of intertwiners. They
permute factors in $\mu_s \circ \mu_\pi $.
To state their properties we
will use the usual permutation $P$,
$$ P: W_\mu \o  W_\nu \mapsto W_\nu \o W_\mu $$
\begin{defn}
{\em (statistics elements)} \label{statEl}
    A family $\R(\mu, \nu)$ which assigns
    a unitary element in $\End(W_\nu) \o \End(W_\mu) \o \K$ to every
    pair $\mu,\nu$ of amplimorphisms
                   is called a family of statistics elements
    if $\hat \R(\mu,\nu) = P \R(\mu,\nu)$ has the following
    properties
\begin{enumerate}
\item   $\hat \R (\mu_1, \mu_2)$ furnishes an intertwiner between
        $ \mu_1 \circ \mu_2$ and $\mu_2 \circ \mu_2$, i.e.
    \be
       \hat \R (\mu_1, \mu_2) \mu_1 \circ \mu_2 (\xi) =
        \mu_2 \circ \mu_1(\xi) \hat \R (\mu_1 , \mu_2)  \ \ ,
    \ee
\item  if $T_i, i= 1,2$ are two intertwiners with the property
       $T_i \nu_i = \mu_i T_i$ then
  \be
     \hat \R (\mu_1, \mu_2) \mu_1(T_2)T_1 = T_2 \nu_2 (T_1)
     \hat \R(\nu_1, \nu_2) \label{TR}
  \ee
\item elements $\hat \R(\mu,\nu)$ satisfy the following
      {\em generalized quasitriangularity} relations
  \ba \label{Drinfeld}
      \hat \R (\mu_1 \circ \mu_2, \nu)&=& \hat \R(\mu_1, \nu)
     \mu_1 (\hat \R(\mu_2, \nu))  \\[1mm]
 \hat \R (\nu , \mu_1 \circ \mu_2 ) &=&
     \mu_1(\hat \R (\nu, \mu_2)) \hat \R(\nu,\mu_1) \ea
\item $\hat \R(\mu,\nu)$ is trivial, if one of its arguments
     is trivial, i.e.
  \be
\hat \R (id_\K, \mu) = \hat \R (\mu, id_\K) = 1_{W_\mu} \o e.
  \ee
\end{enumerate}
\end{defn}
{\bf Remark:} $3.$ are basically Drinfelds
quasitriangularity-relations \cite{Dri2} for
R-matrices which imply the Yang Baxter  equations. They
can be restated with the help of reassociators.
{}From
\ba
\hat R (\mu_\pi \circ \mu_\s, \nu) \vp (\mu_\pi, \mu_\s) & = &
     \nu(\vp(\mu_\pi, \mu_\s)) \hat \R (\mu_{\s \bo \pi}, \nu) \\
\hat R (\nu, \mu_\pi \circ \mu_\s)\nu( \vp (\mu_\pi, \mu_\s)) & = &
     \vp(\mu_\pi, \mu_\s) \hat \R (\nu ,\mu_{\s \bo \pi})
\ea
it follows that
\ba
 \nu(\vp(\mu_\pi, \mu_\s)) \hat \R (\mu_{\s \bo \pi}, \nu)
  \vp^{-1} (\mu_\pi,\mu_\s) & = & \hat \R(\mu_\pi,\nu)
  \mu_\pi(\hat R (\mu_\s, \nu))\ , \\
  \vp(\mu_\pi, \mu_\s) \hat \R (\nu ,\mu_{\s \bo \pi})
  \nu( \vp^{-1} (\mu_\pi, \mu_\s)) & = &
     \mu_\pi (\hat \R (\nu, \mu_\s)) \hat \R(\nu,\mu_\pi) \ .
\ea
In representation theory, the notion of conjugate representation is
important. This translates into the language of amplimorphisms as
follows
\begin{defn}
{\em (conjugation)}
The family $\{\mu_\pi\}$ of amplimorphisms admits a conjugation if there
is                                  for every $\mu$
an amplimorphism $\bar \mu$ and an intertwiner $g(\mu, \bar \mu)
\in \End(W_{\bar \mu} \o W_\mu, {\bf C}) \o \K$ such that
$$ g(\mu , \bar \mu) \mu \circ \bmu (\xi) = \xi g(\mu,\bmu) $$
\end{defn}
We introduce the notation $\chi_\mu \equiv g(\mu,\bmu) \mu
(g^*(\bmu,\mu))$. It has the property
$$ \mu(\xi) \chi_\mu = \chi_\mu \mu(\xi) $$
Sometimes it is convenient to admit the possibility that the
conjugate amplimorphism $\bar \mu $ and the intertwiner $g(\bar\mu ,\mu
 )$ exist, but $\bar \mu $ does not belong to the family which is
associated with representations $\pi \in \Rep$.

\subsection{Differential calculus}

We assume that a family of amplimorphisms is given which
generates all the representations $\pi \in \Rep $ of $\K$
in the sense of definition \ref{Fam}. We assume in addition that
statistics elements are defined, that a reassociator exists, and that
the family of amplimorphisms admits conjugation.

We construct a differential algebra. Here we consider only a special
case. We assume that there exists a special amplimorphism $\mu^f$,
such that all the representations in $\Rep $ can be constructed as
subrepresentations of $\pi_0\circ\mu^f\circ ... \circ \mu^f$, i.e.
as subrepresentations of products of representations
$$ \pi_f =\pi_0 \circ \mu^f \ . $$
The construction can be generalizes to situations with several
fundamental representations. From now on we will write
$\mu $ in place of $\mu^f$.

By assumption, $ \R (\mu , \mu )$ is unitary
and $P^2=id$. Therefore $\hat \R$ can be diagonalized and admits a
spectral decomposition
$$ \hat \R(\mu, \mu) = \sum_{\a = 0}^M \gamma_\a E_\a \ \ \ , \ \ \
   \hat \R(\bmu,\bmu) = \sum_{\a = 0}^M \c_\a \bar E_\a $$
where $\c_a$ are the eigenvalues of $\hat \R (\mu ,\mu )$.
We define
$$ \Pi_\c = \prod_{\c_\a \neq \c} (\hat \R(\mu,\mu) -\c_\a)
            (\c - \c_\a)^{-1}\ \ \ , \ \ \
    \bar \Pi_\c = \prod_{\c_\a \neq \c} (\hat \R(\bmu,\bmu) -\c_\a)
            (\c - \c_\a)^{-1}\ \ . $$
In the following we assume that one of the eigenvalues
$\c \in \{ \c_\a \}$ has been
singled out.

For the sake of clarity, let us introduce a basis
$\{  e^a      \} , ( a     =1...n_f)$ in $W_{\pi^f}$.
This is not necessary, but it
will facilitate comparison with the classical calculus.

We construct a ${\bf Z}$-graded algebra.
\be\Om = \sum_{n\geq 0}\Om_n  \ee
$\Om_0$ is a substitute for the space of functions on a manifold, and
$\Om_n$ substitutes for the space of n-forms.
Elements of $\Om_n$ are said to be of degree $n$.
\begin{defn}{\em (Differential Algebra)}\label{DC1}
The algebra $\Om $ is generated by $\K$, generators $Z_a $
of degree $deg(Z_a)=0$ and generators
$\Theta_a$, $a=1...n_f$ of degree $1$,
                                  subject to the following
relations  for $Z=\sum_a Z_a e^a$, $\Theta  =\sum_a \Theta_ae^a \ .$
\ba
\xi Z = Z \mu(\xi) \ \ \ & , & \ \ \ \xi \Th = \Th \mu(\xi)\ \
\mbox{ for all }\xi \in \K \\
ZZ \Pi_\c = 0 \ \ \ \ & , & \ \ \ \Th \Th \hat \R(\mu,\mu )= \c
                 \Th \Th    \label{BRAID1} \\
Z \Th \hat \R (\mu,\mu) & = & - \c \Th Z    \label{BRAID2}
\ea
$\Om $ is made into a differential algebra by adjoining to it an
operator $d$ subject to the relations
\ba \xi d &=& d \xi \ \ \mbox{for all }\xi \in \K
 \ \                       \mbox{ (invariance)}\\
d^2 &=& 0 \\
dZ &=& \Theta + Zd \ \ \ \ \ \ \ \ \ \ \ \ \mbox{ (Leibniz rule)} \\
d\Theta &=& -\Theta d \ea
\end{defn}
$Z_{a}$ substitute for coordinate functions, and $\Theta_a$
substitute for differentials.\\[2mm]
{\sc Proof} that the definition is meaningful: \\[2mm]
Consistency of the braid relations (\ref{BRAID1},\ref{BRAID2})
follows from the Drinfeld relations, definition \ref{statEl}
in the standard way.\\[2mm]
Consistency of $d$: It must be verified that
the relation $dZ=\Theta + Zd$ is consistent with $ZZ\Pi_\c=0$. We
move $d$ through $ZZ\Pi_\c$.
\ba
d(ZZ\Pi_\c) & = & \Th Z \Pi_\c + ZdZ \Pi_\c = \Th Z \Pi_\c +
Z \Th \Pi_\c + ZZd\Pi_\c \\
& = & Z\Th [1 - \frac{1}{\c} \hat R(\mu,\mu)] \Pi_\c = 0
\ea
Consistency of the other relations with $d$ is obvious q.e.d.

In the classical calculus, $d $ can be constructed from partial
differential operators $\pl_{\dot a} $ and differentials
$\theta_a= dz_a$. This is also possible here.
Our $\Theta_a$ substitutes for the operator
of multiplication
with $\theta_a$, and similarly for $Z_a$.

We introduce a basis $\{ e^{\dot{b}}\}$ in $W_{\bar \pi^f}$,
$\bar \pi^f = \pi_0\circ \bar \mu  $.
\begin{defn}{\em (partial derivatives)}\label{PC}
We adjoin partial derivatives $\pl = \sum_{\dot{b}}\pl_{\dot{b}}
e^{\dot{b}}$ to $\Om $. They are $\K$-covariant in the sense that
\be \xi \pl = \pl \bar \mu (\xi ) , \ee
and are                 subject to the relations
\ba  \label{plRelations}
\pl \pl \bar \Pi_\c = 0 \ \ \ & , & \ \ \  \xi \pl = \pl \bmu(\xi)\\
\Th \pl & = & -\c \pl \Th \hat \R (\bmu,\mu) \\
\pl Z &= & g(\mu,\bmu) - \c Z \pl \hat R (\mu,\bmu)
\ea
\end{defn}
\begin{theo}\label{MainTheorem}
   {\em (Consistency of the differential calculus)}
The relations (\ref{plRelations}) for the partial derivatives
are consistent with the relations in $\Om$.
If we define $d$ in terms of differentials $\Theta $ and
partial derivatives $\pl $ by
$$ d = \Th \pl g^*(\bmu,\mu) \chi^{-1}_\mu ,$$
then $d$ has the properties stated in definition \ref{DC1}; in
particular it satisfies $d^2=0$ and the Leibniz rule.
\end{theo}
The proof requires a number of consistency checks. It will be relegated
to Appendix A.
\section{Examples}

This section is devoted entirely to examples of the structures
introduced above. We start with the simplest commutative case.
Then the above construction will be applied to (quasi-) Hopf-algebras.
In this case we recover the standard calculus
on (quasi-) quantum planes \cite{WeZu,MSVI}.
Algebraic quantum field theory
provides examples where the amplimorphisms are not obtained
from a co-product. This finally leads to quantum field planes
as new examples of non-commutative differential geometries.

\setcounter{equation}{0}

\subsection{Calculus on ${\bf C}^N$}
\def\GL{GL(N,{\bf C})}
The algebra of polynomial (or holomorphic)
functions of $N$ complex variables
is generated by coordinate functions $z_a$, $a=1...N$. We denote by
$Z_a $ the operator of multiplication by $z_a$. The algebra of forms
is obtained by adjoining $\theta_a = dz_a$, $(a=1...N)$. We denote by
$\Theta_a$ the operator of multiplication by $\theta_a$.

The group $\GL $  can act. It possesses a trivial
1-dimensional representation $\pi_0$.
$\pi_0(\xi )=1$ for $\xi \in \GL $.
It has also a defining $N$-dimensional
representation $$\pi^f(\xi )= \xi . $$

We denote by $\Rep $ the set of all those
irreducible representations of $\GL $ by matrices $\pi (\xi )$ which
appear as subrepresentations of products of defining representations.
They act in representation
spaces
$$W_\pi = {\bf C}^{n_\pi} . $$
All these representations  extend to representations
of the group algebra $\K = {\bf C}GL(N,{\bf C})$.

The conjugate $\bar \pi $ of the fundamental representation is defined
by
$$ \bar \pi (\xi )= ^t\xi^{-1} $$
It does not belong to $\Rep$, but we consider it anyway.\\[2mm]
With every  representation  $\pi \in \Rep $ we associate an
amplimorphism
\ba \mu_\pi  :  \K \mapsto End(W_\pi)\o \K &=& Mat_{n_\pi}(\K )\\
     \mu_\pi      (\xi ) &=& \pi     (\xi )\o\xi , \ \ (\xi \in \GL )\ea
It follows that
\ba    \pi_0\circ \mu_\pi (\xi )&=& \pi(\xi )\pi_0(\xi) \\ &=&
    \pi (\xi ) \ \mbox{for } \xi \in GL(N,{\bf C}) \ . \ea
The amplimorphism $\mu = \mu^f$ is associated with the N-dimensional
fundamental representation, $\bar \mu $ with its  conjugate.

The intertwiner between $\mu \circ \bar \mu $ and the identity
morphism                                      has the form
\ba  g(\mu ,\bar \mu )&=&\tilde g(\mu , \bar \mu)e \nn \\
     \tilde g_{ab}(\mu ,\bar \mu)&=& \delta_{ab} \nn
\ea
It follows that
$  \tilde g_{ab}(\mu , \bar \mu) ^t\xi^{-1}_{ac}\xi_{bd}=
\tilde g_{cd}(\mu ,\bar \mu)$ as required.

Finally we define $\R (\mu ,\mu )= id \o id $, so that
$\hat \R $ is the permutation operator.
\be \hat \R(\mu , \mu )= P, \ee
i.e. $\hat \R(\mu , \mu )_{ab,cd}= \delta_{ad}\delta_{bc}.$
It has two eigenvalues
\be \gamma_0 = +1 \ \ , \ \ \gamma = \gamma_1=-1 \ee
The reassociator is trivial, $\vp (\mu_\pi, \mu_\s)= id\o e$.

It is now readily verified that the formulae of the differential
calculus as set out in definitions \ref{DC1},\ref{PC}
                                                reproduce the classical
calculus.

The algebra $\Om $ of section \ref{DC} is generated by
$\K = {\bf C}\GL$ and generators $Z_a$, $\Theta_a$. Here it is the
algebra of polynomial or holomorphic ${\bf C}\GL $-valued functions
and forms of $N$ complex variables $z_1...z_N$. For convenience
we use multiplication operators, but this makes no difference
in cases with a trivial reassociator. The braid relations
\ref{BRAID1},\ref{BRAID2} become
\ba Z_aZ_b = + Z_bZ_a \ \ &,& \Theta_a\Theta_b=-\Theta_b\Theta_c
\label{Pbraid1}\\
       Z_a\Theta_b &=& \Theta_bZ_a \label{Pbraid2}\\
       dZ_a &=& \Theta_a + Z_a d \ . \ea

Both generators $Z_a$ and $\Theta_a $ transform according to the
defining  representation of $GL(n,{\bf C})$, viz.
\ba \xi Z_a = Z_b\mu_{ba}(\xi ) = Z_b\pi_{ba}(\xi )\xi , \\
\xi \Theta_a = \Theta_b\mu_{ba}(\xi ) = \Theta_b\pi_{ba}(\xi )\xi \ea
Therefore exterior differentiation $d$ commutes with the action of
$\GL $.

The partial derivatives commute among themselves and with $\Theta_a$.
They                    transform according to the conjugate
of the defining representation.
\ba \pl_{\dot a}\pl_{\dot b}= \pl_{\dot b}\pl_{\dot a } \ \ \ &,&
\xi \pl_{\dot a} = \pl_{\dot b}\bar \mu_{ba}(\xi ) =
\pl_{\dot b} (^t\xi^{-1})_{ba}\xi   \\
\Theta_a \pl_{\dot b} &=&   \pl_{\dot b}\Theta_a \\
\pl_{\dot a}Z_b &=& \delta_{\dot a b} e + Z_b\pl_{\dot a} \\
d &=& \Theta_a \pl_{\dot b} \delta_{a\dot b} . \ea
\subsection{Quantum planes}\label{QPlanes}
In the quantum planes and their generalizations, the symmetry group
$\GL $ or $SU(N)$ of the classical calculus is replaced by a
quantum symmetry. In the most general case it can be a weak
quasitriangular quasi Hopf *-algebra $\G $ \cite{MSIII}.

Throughout this subsection, let \\[2mm]
$\K = \G$ - semisimple *-algebra with unit e \\
$\Rep$ - set of all unitary representations $\t$ of $\G$\\
$\pi_0 = \e$ - the co-unit of $\G$ \\
$W_\t = V_\t$ - carrier space of the representation $\t$ \\[2mm]
The counit $\epsilon $ defines a 1-dimensional representation of $\G$.
i.e. a homomorphism of $\G $ into ${\bf C}$.
$\G $ comes equipped with a comultiplication
\be \Delta : \G \mapsto \G \o \G .\ee
$\Delta $ is a not necessarily unit-preserving *-homomorphism of
algebras. If $\Delta (e)\not=e\o e $ we speak of truncation.
It is required that
\be (id \o \epsilon )\Delta = id = (\epsilon \o id) \D \label{epsDel}\ee
$\Delta $ is required to be quasiassociative in the following sense.
There exists an element $\vp \in \G \o \G \o \G $ which possesses
a quasi-inverse $\vp^{-1}$ such that
\be \vp (\Delta \o id )\Delta (\xi )=
(id \o \Delta )\Delta (\xi )\vp \ \ (\xi \in \G ). \ee
The quasi-inverse property is
\be \vp \vp^{-1}= (id \o \Delta )\Delta (e) \ \ , \ \
    \vp^{-1}\vp = (\Delta \o id )\Delta (e) \ . \ee
The coproduct is required to be
braid commutative in the following sense.\\[2mm]
          Let $\Delta^{\prime}(\xi ) $ differ from $\Delta (\xi )$
by interchange of factors. There exists an element $R\in \G \o \G $
with quasi-inverse $R^{-1}$ such that
\be  R \Delta (\xi )= \Delta^{\prime}(\xi ) \ . \ee
The quasi-inverse property states that
\be RR^{-1}= \Delta^{\prime}(e)\ \ , \ \
    R^{-1}R = \Delta (e)\ . \ee
$R$ and $\vp $ must satisfy a number of consistency relations -
Drinfelds hexagon and pentagon identities. There is also an antipode
$\S $ with certain properties. It is an anti-automorphism of $\G $.

The representation theory of the quantum symmetry $\G $ can be
rephrased in the language of amplimorphisms by defining
amplimorphisms and their intertwiners as follows \cite{Vec}.
\ba
\mu_\t(\xi) & = & (\t \o id) \D (\xi) \ \ \mbox{ for all }
  \ \ \xi \in \G \\[1mm] \nn \ \ , \\
  \vp(\mu_\t, \mu_\s) & = & (\s \o \t \o id) (\vp)\ \ , \nn \\[1mm]
   \R ( \mu_\t, \mu_\s) & = & (\s \o \t \o id )
   (\vp_{213} R_{12} \vp^{-1})  \ \ . \nn
\ea
We use the standard notations: If $\vp_{123}= \sum_\s
\vp_\s^1\o\vp_\s^2\o\vp_\s^3$ then
                               $$  \vp_{s(1)s(2)s(3)}=
 \sum_{\s }\vp_\s^{s^{-1}(1)}\o
            \vp_\s^{s^{-1}(2)}\o\vp_\s^{s^{-1}(3)}$$
etc.
The properties of $\vp, \R$ can be checked. The equations in the
remark after the definition of statistics elements are
equivalent to the following equations for $R, \vp$.
\ba
[(id \o id \o \D)(\vp)]_{2314} (id \o \D \o id) (\R) (e \o \vp^{-1})
& = & \R_{134} (id \o id \o \D) \R \nn \\
\vp_{124} (\D \o id \o id)(\R)(id \o id \o \D)(\vp^{-1}) & = &
 [(id \o id \o \D)(\R)]_{1324} \R_{234} \nn
\ea
These relations are in turn equivalent to Drinfelds hexagon and
pentagon equations. \\[2mm]
The conjugate amplimorphism is defined with the help of the
antipode,
$\bmu_\t = \mu_{\tt}$ with $\tt(\xi) = \ ^t\t(\S^{-1}(\xi))$.
The intertwiner $g(\bmu ,\mu )$ exists \cite{Vec,MSIII}.
\\[2mm]
The definition of the product of representations with the help of
amplimorphisms is equivalent to the usual definition of
tensor products of representations of quantum symmetries
\be
     (\t_2 \bo \t_1)(\xi) =
     (\t_2 \o \t_1)(\Delta (\xi )) \ . \label{prodTau} \ee
We assume  that all the irreducible representations of $\G$ can
be obtained by reducing products of one fundamental representation
$\tau^f = \tau $ of $\G$ of dimension $n_f$. The associated
amplimorphism will be denoted by $\mu^f = \mu $.

The algebra $\Om $ of section \ref{DC} will be generated by $\G $
together with
$n_f$ generators $Z_a$ of degree $0$ and $n_f$ generators $\Theta_a$
of degree $1$ as before. This algebra substitutes for an algebra
of $\G$-valued functions and forms of $n_f$ noncommuting variables
$z_a$. The algebra is only quasi-associative if $\G $ is not
coassociative, i.e. if $\vp \not= e\o e\o e$. In the
quasiassociative case it is essential to have $\G $ as a part of the
algebra to start with. $\G $ can be factored out also in this case,
but this involves a nontrivial construction, see \cite{MSVI}.

If $\tau^f \o \tau^f(R) $ possesses only two different eigenvalues
$\gamma_0 $ and $\gamma $, then the construction of section \ref{DC}
reproduces the well known quantum planes and their quasi-associative
generalizations \cite{MSVI}.
                 The $ZZ$ braid relations reduce in this case to
\be ZZ\hat \R (\mu , \mu )=\gamma_0 ZZ \ . \ee
In the coassociative case, $\R (\mu , \mu )=(\tau \o \tau )(R)\o e $,
i.e. it is a numerical matrix. So we get the  standard quantum
planes.
\subsection{Algebraic quantum field theory}\label{AQFT}
In special relativistic local quantum field theory there is a
*-algebra $\A $ of observables. For convenience,
                                one considers complex linear
combinations of bounded self adjoint observables as elements of
$\A $. There are subalgebras $\A (\O )$ which contain observables
  that can be measured in bounded contractible domains $\O $ of
space time. Locality says that
            observables in relatively spacelike domains commute.
There is also a Hamiltonian $H$ affiliated with $\A $. In a positive
energy representation it has nonnegative spectrum.

Typically, the algebra $\A $ has a number of inequivalent irreducible
positive energy representations $\pi $. They are called
superselection sectors \cite{WWW}.

Among these representations is the vacuum representation $\pi_0$. It
acts on a Hilbert space $\H_0$ which contains the ground state of $H$.
This sector is called the vacuum sector.

One starts from
general assumptions of algebraic field theory.
These  assumptions are  fulfilled
in conformal field theories, provided one completes $\A (\O )$ to von
Neumann algebras (acting in $\H^0$), and the global algebra $\A $ is
constructed from the local algebras $\A (\O )$ in an appropriate way
\cite{Baumg,Fre3}. It follows that there
exist localizable unital *-endomorphisms
\be
\rho^J: \A \mapsto \A
\ee
such that all positive energy representations $\pi^J$
                                              can be obtained from
the vacuum representation by composition with these morphisms
\be
 \pi^J=\pi^0\circ \rho^J \ \ \mbox{i.e.} \ \ \pi^J(A)=\pi^0(\rho^J(A)).
\ee
By definition, unital *-endomorphisms of $\A $ are ${\bf C}$-linear
maps subject to the conditions
\ba
                                 \label{morphism}
    \rho (AB) & = & \rho (A) \rho (B) \ \ ,\nn \\
\rho (A^*) &=& \rho (A)^* \ \ ,\nn\\
\rho ({\bf 1}) &=&  {\bf 1}\ \ .
\ea

Localizability means the following. To every bounded space time domain
$\O $ there exists a unitary $u\in \A $ such that the morphism
\be \rho^J_u (A)= u^{\ast }\rho^J (A)u ,\ee
is localized in $\O $. This means that
\be \rho^J_u (A)=A \ \ \mbox{if } \ \ A\in \A (\O^{\prime }) \ee
and if $\O^{\prime }$ is relatively spacelike to $\O $.

Using these morphisms, a  product of irreducible positive
energy representations of $\A $ can be defined as
\be \pi^I\bo    \pi^J = \pi^0\circ \rho^I\circ\rho^J \ . \ee
These products will decompose into irreducibles according to
\be \pi^I\bo    \pi^J = \sum_K \pi^KN_K^{IJ}  \label{posEdec}
  \ee
with multiplicities $N_K^{IJ}$.
It follows from the standard assumptions of algebraic field theory
that the observable algebra determines  statitics operators
$\epsilon(\rho^J,\rho^K)\in \A $ which enjoy
standard properties \cite{Mebk}.

One defines   spaces of intertwiners between morphisms,
$$T \in
                   \T(\rho_1,\rho_2) \subset \A \ \ \ \mbox{ iff }
 T \rho_1 (A) = \rho_2 (A) T \ \ \ \mbox{ for all} \ \ A \in \A\ \ .$$
For every pair of localizable endomorphisms $\rho_i,i=1,                2$
there is a unitary local intertwiner, the
$$ \mbox {statistics operator } \ \ \
                                      \e(\rho_1,\rho_2) \in
\T(\rho_1\circ \rho_2, \rho_2 \circ \rho_1)$$
Explicitly, its intertwining property is
$$
                                      \e(\rho_1,\rho_2)
\rho_1\circ \rho_2 (A) =
\rho_2\circ \rho_1 (A)
                                      \e(\rho_1,\rho_2)
$$
           The collection of statistics operators
in uniquely determined by the following
equations (a detailed proof can be found in \cite{Mebk})
\ba
\e(\rho_1,\rho_2) \rho_1 (T_2 ) T_1 & = & T_2 \s_2 (T_1) \e(\s_1,\s_2)
\ \ \ \ \mbox{ for all } \ \ \ T_i \in \T(\s_i,\rho_i) \nn \ \ ,\\[2mm]
\e(\rho_1 \circ \rho_2, \s ) = \e(\rho_1,\s ) \rho_1(\e (\rho_2,\s ))
\ \ \ &  & \ \ \ \e(\s,\rho_1 \circ \rho_2) = \rho_1 (\e (\s,\rho_2))
\e(\s, \rho_1) \label{epsprop}\ \ ,\\[2mm]
\e (\rho_1,\rho_2) &=& 1 \ \ \ \mbox{ whenever } \ \ \ \rho_1  > \rho_2
\ \ .\nn \ea
In the last row, $\rho_1 > \rho_2$ if $\rho_1 $ and $\rho_2$
are localized in relatively spacelike domains, and if these domains
are ordered, when they can be ordered. This last property
will not be  used here. In chiral conformal field theories on
                         the circle, ordering
is left or right, after a reference point on the circle has been
chosen.
Trivialization for $\rho_1< \rho_2$ would give rise to the
{\em opposite statistics operator}
 $\e(\rho_2,\rho_1)^*$.

Here we restrict our attention to sectors with finite statistics.
For such sectors, the sum in the decomposition (\ref{posEdec})
of the product or representations is finite. This assumption is used
in the construction of field operators in the next section.

All this fits in with the setup described at the beginning of
section \ref{DC} if we make the following identifications
\\[2mm]
$\K = \A$ - observable algebra of a QFT satisfying standard
assumptions \\
$\Rep $ - set of positive energy
                  representations with finite statistics\\
$\pi_0$ vacuum representation of $\A$ \\
$W_\pi = {\bf C}$,
\\
$\R(\rho ,\sigma )
 = \e ^* (\sigma , \rho )$ ( one could also take $\e(\rho , \sigma)$)
\\[2mm]
      The          amplimorphisms $\mu$ are the endomorphisms $\rho^J$
\\[2mm]
Every sector with finite statistics has a unique conjugate. We
denote by $\bar \rho $ the morphism which generates it.
It can be shown that the intertwiner $g(\brho ,\rho )$ exists for
sectors of finite statistics and has the form
$$g(\brho,\rho) = \sqrt{d_\rho} R^*_\rho .$$
where                   $R_\mu$  is an isometry with property
   $ \brho\circ \rho(\xi) R_\rho = R_\rho \xi $.  \\
$d_\rho > 0 $ is called the statistics  of the
   sector generated by $ \rho $.\\
{\bf Remark:} phases in $R_\rho$ can be adjusted such that
$ \chi_\rho = \pm 1$, where the sign depends on the sector but
$+$ for all sectors which are not selfconjugate \cite{FRS2}. \\[2mm]
The reassociator is trivial. \\[2mm]
This setup is not suitable yet to construct an interesting differential
calculus because the spaces $W_\pi $ are trivial.
But this changes when the gauge symmetry of the theory is taken into
account.
\subsection{Quantum field planes}\label{QFPlanes}
A gauge symmetry in quantum mechanics is a symmetry which leaves the
observables invariant. Here we are interested in gauge symmetries of
first kind (i.e. global ones), to start with. According to the
principles of gauge theories and of general relativity, these
gauge symmetries should become local symmetries (gauge symmetries of
second kind), when the local frames are glued together in a covariant
fashion.

Let us start from an algebra of observables in a local quantum field
theory and its superselection sectors as described in the last section.

We assume that there exists a quantum symmetry $\G $ which encodes
the selection rules \\(\ref{posEdec}). In detail this means that there
should exist a semisimple
               bi-*-algebra $\G $ such that the irreducible
representations $\tau^J$ of $\G $ are in bijective correspondence with
the superselection sectors $\pi_J$, and the selection rules match. That
is
\be \tau^I\bo    \tau^J = \sum_K \tau^KN_K^{IJ}  \label{taudec}    \ee
with the same multiplicities $N_K^{IJ}$ as in (\ref{posEdec}).

This requirement makes sense. By definition, a bi-*-algebra $\G$ is
a *-algebra with unit which is equipped with a coproduct
$\Delta $ and a counit $\epsilon $. They are *-homomorphisms of
$\G $ into $\G \o \G $ resp. ${\bf C}$, and are subject to
the relation (\ref{epsDel}). The product $\bo $ of representations
$\tau $ of $\G$ is therefore defined by eq.(\ref{prodTau}) and
admits a decomposition into irreducibles by semisimplicity.

It was shown by one of us \cite{VSDiss} that one can find the
necessary intertwiners to make $\G $ into a weak quasitriangular
quasi Hopf algebra as described in subsection \ref{QPlanes}.
The input for this construction is furnished by the
algebra of observables, its morphisms and their intertwiners as
described in subsection \ref{AQFT}. Using this, the amplimorphisms
for $\G $ and their intertwiners can be constructed as in subsection
\ref{QPlanes}. We write $\mu^J_\G$ for the amplimorphism of $\G $ which
is associated with its representation $\tau^J$, etc.

We take the tensor product of both structures. In the following, letters
$a$ will stand for elements of $A$ and $\xi $ for elements of $\G $.
By definition, $\rho^J(a)\in \A$, and $\epsilon (\xi )\in {\bf C}$. Let
\\[2mm]
$\K = \G \o \A \ $,\\
$\Rep$ - representations $\pi^J(a\xi) = \pi^J(a)\o \tau^J(\xi )$,\\
$\pi_0 (\a\xi )= \epsilon (\xi )\pi_0(a)$. It acts in $\H_0$.\\
$W_{\pi^J}\equiv W_J
          = W_{\tau^J}$ - carrier space of the representation $\tau^J$
of $\G $ \\[2mm]
The amplimorphisms and their intertwiners become
\ba
\mu^J(a\xi ) & = &  \mu^J_\G(\xi )\rho^J(a) \in \End(W_J)\o \K \ \ , \nn\\
\vp(\mu^J,\mu^K) & = & \vp_\G(\mu^J_\G , \mu^K_\G ) \ \ , \nn \\
\R(\mu^J,\mu^K) & = &
 \R_\G(\mu_\G^J,\mu_G^K) \epsilon^{\ast }(\rho^K,\rho^J) \ \ , \nn
\ea
and similarly for $g(\bar \mu^J, \mu^J) $.

These quantities satisfy the requirements laid down at the beginning
of section \ref{DC}, because the factors from $\G $ and from $\A $ do.

Now we can introduce the differential calculus. We assume for
simplicity that all the representations $\tau^J$ of $\G $ are
contained in products of a single fundamental representation
$\tau^f=\tau$ of dimension $n_f$.
       The corresponding endomorphism of $\A $ is denoted by
$\rho^f=\rho$.

The algebra $\Om $ is generated by generators $Z_\a$ of degree $0$
and $\Theta_\a$ of degree $1$, $\a = 1... n_f$.
The multiplication operators $\Theta_\a$ will be called
{\it protofields\/} for reasons that will become clear in the next
section.

Their relations  are as stated in section \ref{DC}.
Let us spell out some of them. Covariance under $\G \o \A$ yields
\ba
    a \Theta_\a &=&
         \Theta_{\a }\rho^f(a) \ \ \ (a \in \A ) \label{aThComm} \\
    \xi \Theta_\a &=& \sum \Theta_{\b }\tau_{\b \a }(\xi^1)\xi^2 \ \ \
(\xi \in \G ) \ , \ea
if $\Delta (\xi )= \sum \xi^1 \o \xi^2 $. The same relations hold
with $Z $ substituted for $\Theta$.

Similarly the braid relations (\ref{BRAID1}) for $\Theta $ read
\ba \Theta_\a \Theta_\b \R^\G_{\b \a , \k \delta }\epsilon^{\ast }
(\rho^f , \rho^f) &=& \gamma \Theta_\k \Theta_{\delta } \\
\R^\G &=& (\tau^f \o \tau^f \o id)(\vp_{213}R_{12}\vp^{-1})\ .\ea
$\R^\G $ is constructed from the $R$-element of $\G \o \G$ and the
reassociator for $\G $, as in subsection \ref{QPlanes}. If
$\G $ is coassociative then $\R^\G_{\b \a ,\c \delta }$ is a number,
otherwise it is an element of $\G $. $\epsilon^{\ast }(...) $ is an
$a$-number.                          $\gamma $ is a complex number
(of modulus 1).\\[2mm]
                If $\R^\G $ has only two distinct eigenvalues then
the braid relations for $Z$ look the same as for $\Theta $,
   with another complex
factor $\gamma_0$.
The $Z\Theta $-braid relations look the same as the $\Theta \Theta $
braid relations except for an overall -sign.

Let us note that all these relations are very much the same as in a
quantum plane, except for the appearance of $a$-numbers in place of
c-numbers. Note however that $a$-numbers do not commute with
generators $\Theta_\a $ and $Z_\a $. Instead one has the
commutation relations \\(\ref{aThComm}), and the same with $ Z $
substituting for $\Theta $.

It is appropriate to call the number $n_f$ of generators $Z_a$ the
{\it dimension\/} of the quantum space time. It is determined by the
symmetry $\G $ and is independent of the dimension of the classical
space time in the local Lorentz frames.
\section{Differential geometry}
Let $A\in \Om_1 $. It substitutes for
                  a $\G\o \A$-valued 1-form. The covariant exterior
derivative
\be D=d+A \ee
transforms covariantly under invertible elements $\Xi \in \F=\Om_0$.
$\Xi $ substitutes for a $\G \o \A$-valued function.
\ba D\Xi &=&\Xi D^{\prime }\\
D^{\prime }&=&d+A^{\prime}\ , \\
A^{\prime} &=& \Xi^{-1}A \Xi + \Xi^{-1}[d,\Xi ] \ . \ea
The field strength $F\in \Om_2$ is defined in the costumary way,
\be F=D^2 = \{ d ,A \} + AA \ . \ee

\section{Operator product expansions}
\setcounter{equation}{0}

It is possible to impose additional relations in the algebra $\Om$.
They are the operator product expansions of quantum field theory.
They are more restrictive than the braid relations alone.

The algebra generated by $\G \o \A $ and the protofields $\Theta_\a$
becomes the algebra of field operators of the  local field theory
with observable algebra $\A$ and gauge symmetry of first kind $\G $,
upon imposing operator product expansions for the $\Theta$'s. More
precisely $\Phi $ includes besides the quantum fields also the
representation operators  of the elements $\xi $ of the symmetry
$\G $; we use the same symbol $\xi $ for them.

The operator product expansions look as follows.
The algebra $\Phi $ is the $\A$-linear span of basis vectors
$\Theta^J_\a $, $\a = 1...n_J=$dimension of representation $\tau^J$ of
$\G $. They are $\K $-covariant in the sense that
\be  a\xi \Theta^J = \Theta^J \mu^J(a\xi ) \ . \label{axiThComm} \ee
       There exist\\[2mm]
{\bf intertwiners:} $C \vvert{K}{I}{J} \in \End (W_J \o W_I, W_K)
\o \K$ with
$$ C \vvert{K}{I}{J} \mu^I \circ \mu^J (\xi) =
\mu^K (\xi) C \vvert {K}{I}{J}.$$
One may impose  orthonormality conditions on them. The operator product
expansions are the multiplication law for the basis vectors. They read
\be
\Theta^J \Theta^I = \sum_K \Theta^K C \vvert {K}{I}{J} \ . \label{OPE}
\ee
In general
$ C \vvert {K}{I}{J} $ are matrices with entries in $\G\o\A$. In case
$\G $ is coassociative, the entries are $a$-numbers.

These operator product expansions look very much like those
in topological field theory \cite{topGrav}. The difference is that
the coefficients are $a$-numbers in place of c-numbers.
In fact they are convergent and partially summed up
                            forms of the standard operator
product expansions, as are familiar from conformal field theory.
They are valid in any quantum field theory which satisfies the
standard assumption of algebraic field theory. They were proven in
 \cite{VSDiss}.
\footnote{It was shown long ago that operator product expansions in
conformal field theory are convergent \cite{MacOPE} }
One can use the operator product expansions to construct all the
protofields $\Theta^J$ from products
                          of the fundamental $\Theta$'s.

The operator product expansions will in general not respect the
${\bf Z}$-grading. We assume, however,that they admit a ${\bf Z}_2$
symmetry. This leads to a surviving ${\bf Z}_2$-grading.

\begin{theo}{\em (Consistency of operator product expansions)}
\label{PDCtheo}
The operator product expansions (\ref{OPE}) are consistent with the
other relations in the algebra $\Om$, except for the ${\bf Z}$ grading.
\end{theo}
{\sc Proof}: The consistency of the $\Theta\Theta $ braid relations with
with the operator product expansions is an intrinsic property of the
field algebra $\Phi $ which is well known \cite{VSDiss}\\[2mm]
The consistency of the $Z\Theta $ braid relations with
the operator product expansions must be checked. Multiply the difference
of the right hand side and left hand side of (\ref{OPE}) by $Z$
and push the $Z$'s through the $\Theta$'s. The result must be zero.
This is indeed true as a result of properties  of
$\hat \R$, definition \ref{statEl}.
\ba
& & \mu^L (C\vvert{K}{I}{J} ) \hat \R (\mu^I, \mu^L )
 \mu^I (\hat \R (\mu^J, \mu^L )) \nn \\
& = & \mu^L( C( \vvert{K}{I}{J}  ) \hat \R ( \mu^I \circ
    \mu^J, \mu^L ) \nn \\
& = & \hat \R (\mu^K, \mu^L ) C \vvert{K}{I}{J}
\ea
\subsection*{Comments on local fields}
To explain the relation with the standard lore,
let us first explain what the local field operators are.

                                                   One defines the
extended algebras $\Phi (\O )$ of field operators which are localized in
space time regions $\O $. $\Phi (\O )$    consists of those
elements of $\Phi $ which commute with all observables
$a \in \A(\O^{\prime })$. Herein
                         $\O^{\prime } $ is the spacelike complement of
$\O $ as before.
The qualification "extended" refers to the fact that
$\G \subset \Phi (\O )$ for all $\O $.

Let us suppose that $\Psi^J = \Theta^Ju $ with $u $ a unitary element
of $\A $. Then one can use eq. (\ref{axiThComm}) to
compute
\ba  a\Psi^J    &=&
                   \Psi^J    \rho_u(a) \ ,  \\
\rho_u(a) &=&                  u^{-1}\rho^J(a)u
\ea
Thus,
$\Psi^J \in \F (\O )$ if $\rho_u$ is localized in $\O $, cp.
section \ref{AQFT}.

{}From (\ref{OPE}) operator product
expansions for localized
fields $\Psi^J$ are obtained. One uses the a-covariance relation
\ref{axiThComm} to push factors $u$ through $\Theta $. For further
details see \cite{VSDiss}.

Fields  $\Psi^J(x)$ at a point can in principle
be obtained as limits of fields in $\Phi (\O )$ when $\O $ shrinks to
a point. A simple example is found in \cite{BMT} where this
construction is carried through explicitly.

Let us  briefly recall how the Clebsch Gordan kernels
 $ C \vvert {K}{I}{J} $  are constructed. This will make it clear
why one needs the symmetry $\G $ to construct field operators
which obey operator product expansions on the whole Hilbert space of
physical states.
\\[2mm]
                                          There exist
intertwining operators
$ T^i \vvert {K}{I}{J}$ for $\A $ \cite{FRS,MSI,VSDiss}
                                 . They are used to reduce the
representation $\pi^I \bo \pi^J$ of $\A $. Because the representation
$\pi^K$
         appears in $\pi^I\bo \pi^J $ with multiplicity $N_K^{IJ}$,
these intertwiners are labelled by $i=1...N_K^{IJ}$.

Similarly there are
Clebsch-Gordan kernels
$ C^i_\G \vvert {K}{I}{J}$ for $\G $. By assumption, the representation
$\tau^K$ appears in $\tau^I\bo \tau^J $ with multiplicity
$N_K^{IJ}$. Therefore these Clebsch Gordan kernels are also labelled by
$i = 1 ... N_K^{IJ}$, with the same multiplicities $N_K^{IJ}$.
One may contract over the index $i$ to obtain
\be C \vvert {K}{I}{J}=\sum_i
  C^i_\G\vvert {K}{I}{J} T^i\vvert {K}{I}{J} \ . \ee
\section{Divorce of the Lorentz group}
\setcounter{equation}{0}

When one starts from a 4-dimensional field theory, the symmetry $\G $ is
a compact group\cite{DoRo}. In other cases it substitutes for a compact
internal symmetry group. Here we wish to use it as a substitute
for the Lorentz group. This requires a comment.

\noindent
Classical general
relativity may be regarded as a gauge theory with gauge group $SO(3,1)$.
Note that the local symmetry in global space time is a global
symmetry in local Lorentz frames. Of course, $SO(3,1)$ is noncompact.
But its finite dimensional
representations are tensor products of the "left handed"
and "right handed"    fundamental representations
$(0,\frac{1}{2})$ and $(\frac{1}{2}, 0)$. These representations are
both unitary irreducible representations of $SU(2)$, extended to
$SL(2,\C)$ by analyticity.

In their geometric interpretation of the standard model, Connes and Lott
proposed that space time is double sheeted \cite{ConnesLott}.
                                            It is then natural to
postulate that left handed matter fields live on one sheet, and
right handed matter fields on the other. In this way the Lorentz group
gets divorced into two SU(2) groups. At the level of the fundamental
theory there is no need for a group which accomodates representations
which are tensor products of representations for left handed and for
                                                                 right
handed matter. That comes only at the level of effective theories
at
length scales longer than the distance between the two sheets of space
time.

This suggests that we may replace the Lorentz group by a compact group
like SU(2) or possibly a deformation of it, i.e. by some
weak quasi Hopf algebra \cite{MSIII}.

Since the algebra $\A $ is to become part of the local
                                                 gauge symmetry, there
can  be two algebras $\A^{left } $ and $\A^{right } $ attached to
to the two sheets of space time.

\section{Appendix A: Proof of Theorem 9}
\setcounter{equation}{0}

Consistency of the differential calculus\\[2mm]
(1) check that the constant term in the $\pl-Z$ equations gives no
    new relations.
 \ba
 \pl Z Z \Pi_\c
 & = &  Z \mu (g(\mu,\bmu)) \Pi_\c - \c Z \pl Z
         \mu(\hat R(\mu,\bmu))\Pi_\c \nn \\
 & = &  Z [ \mu(g(\mu,\bmu)) - \c g(\mu,\bmu)
        \mu(\hat \R(\mu,\bmu))]\Pi_\c
        + \c^2 ZZ \hat \R(\mu,\bmu) \mu(\hat R(\mu,\bmu)) \Pi_\c
        \nn
 \ea
We derive two formulas for the intertwiners \\
i) Since $\mu \circ \mu (\xi) \Pi_\c = \Pi_\c \mu \circ \mu(\xi)$
it follows that
$$ \hat \R(\mu, \bmu) \mu (\hat \R (\mu, \bmu))\Pi_\c =
\hat \R(\mu\circ \mu,\bmu)\Pi_\c =
\bmu(\Pi_\c) \hat \R(\mu\circ \mu, \bmu) $$
ii)
\ba
& & \mu(g(\mu,\bmu)) - \c g(\mu,\bmu) \mu(\hat \R (\mu,\bmu)) \nn\\
& = & \mu(g(\mu,\bmu)) - \c g(\mu,\bmu) \mu (\hat \R (\mu,\bmu))
      \hat \R (\mu,\mu) \hat \R^{-1} (\mu,\mu) \nn \\
& = & \mu(g(\mu,\bmu)) - \c g(\mu,\bmu) \hat \R(\mu,\mu \circ \bmu)
      \hat \R^{-1}(\mu,\mu) \nn \\
& = & \mu(g(\mu,\bmu)) - \c \mu(g(\mu,\bmu)) \hat \R^{-1} (\mu,\mu )
\nn \\
& = &  \mu(g(\mu,\bmu)) \hat \R^{-1}(\mu,\mu) [\hat \R(\mu,\mu) - \c ]
\nn \ea
It follows that $\pl ZZ \Pi_\c = 0 $ . \\[2mm]
(2) the relation $dZ = \Th + Zd$.
\ba
dZ & = & \Th \pl g^*(\bmu,\mu)\chi_\mu^{-1} Z\nn \\
& = & \Th g(\mu, \bmu) \mu(g^*(\bmu,\mu))\chi_\mu^{-1} - \c \Th Z \pl
       \hat \R (\mu ,\bmu) \mu(g^*(\bmu,\mu))\chi^{-1}_\mu \nn \\
& = & \Th  + Z \Th \pl
       \bmu(\hat \R(\mu,\mu)) \hat \R(\mu,\bmu) \mu(g^*(\bmu,\mu))
       \chi_\mu^{-1}    \nn
\ea
for the intertwiner we obtain
$$ \bmu(\hat \R(\mu,\mu)) \hat \R(\mu,\bmu) \mu(g^*(\bmu,\mu))
= \hat R( \mu, \bmu\circ \mu) \mu(g^*(\bmu,\mu)) = g^*(\bmu,\mu)$$
so that the Leibniz rule follows. \\[2mm]
(3) To prove $d^2 = 0 $ we use the following lemma \\
{\bf Lemma :} $\bar \Pi_\c \bmu (g^*(\bmu,\mu)) g^*(\bmu,\mu) =
    \bmu \circ \bmu (\Pi_\c) \bmu (g^*(\bmu,\mu))
    g^*(\bmu,\mu) $ \\[1mm]
{\sc Proof:}
\ba
\bmu(\hat \R^* (\bmu,\mu)) g^*(\bmu,\mu) & = &
  \hat \R (\bmu,\bmu) \hat\R^* (\bmu,\bmu) \bmu (\hat \R^*(\bmu,\mu))
   g^* (\bmu,\mu) \nn \\
& = &  \hat \R(\bmu, \bmu) \hat \R^* (\bmu, \bmu \circ \mu)
      g^* (\bmu, \mu) \nn \\
& = &  \hat \R (\bmu, \bmu ) \bmu (g^* (\bmu, \mu )) \nn
\ea
\ba
\bmu(\hat \R(\mu,\mu)) g^* (\bmu,\mu)
& = &
\bmu (\hat \R (\mu, \mu)) \hat \R ^* (\bmu \circ \mu, \mu)
\mu(g^* (\bmu, \mu) ) \nn \\
& = &
\bmu(\hat R (\mu, \mu)) \bmu (\hat \R^* (\mu, \mu))
\hat \R^* (\bmu, \mu) \mu (g^* (\bmu, \mu)) \nn \\
& = &
\hat \R^* (\bmu, \mu) \mu (g^* (\bmu, \mu)) . \nn
\ea
{}From these two formulas we infer
\ba
\hat\R (\bmu , \bmu ) \bmu (g^* (\bmu , \mu )) g^* (\bmu , \mu )
 & = &  \bmu (\hat \R^* (\bmu , \mu ))
                          g^* (\bmu , \mu ) g^* (\bmu ,\mu )
\\
& = & \bmu ( \hat \R^* (\bmu , \mu )) \bmu \circ \mu (g^* (
    \bmu , \mu )) g^* (\bmu , \mu ) \nn \\
& = & \bmu \circ \bmu (\hat \R (\mu, \mu )) \bmu ( g^* (\bmu , \mu ))
      g^*(\bmu, \mu ) \nn
\ea
This proves the lemma.\\[2mm]
Now we are prepared to calculate $d^2$.
\ba
d^2 & = & \Th \pl \Th \pl \bmu \circ \mu( g^* (\bmu, \mu))
             g^*(\bmu,\mu) \chi^2_\mu \nn \\
& = & - \Th \Th \pl \pl \bmu (\hat \R^*(\bmu\circ \mu, \mu))
      \bmu \circ \mu (g^*(\bmu, \mu)) g^* (\bmu,\mu)
      \chi^2_\mu \nn \\
& = & - \Th \Th \pl \pl \bmu (g^* (\bmu, \mu) ) g^* (\bmu, \mu)
       \chi^2_\mu \nn \\
& = & - \Th \Th \pl \pl (1 - \bar \Pi_\c ) \bmu( g^* (\bmu, \mu))
       g^* (\bmu, \mu) \chi^2_\mu \nn \\
& = & - \Th \Th \pl \pl \bmu \circ \bmu(1 - \Pi_\c) \bmu( g^*
      (\bmu, \mu )) g^*( \bmu, \mu) \chi^2 _\mu = 0   \nn
\ea
because $ \Th \Th ( 1 - \Pi _\c) = 0 $ .

\end{document}